\documentclass[conference]{IEEEtran}
\IEEEoverridecommandlockouts
\usepackage{amssymb} 
\usepackage{amsmath} 
\allowdisplaybreaks
\usepackage{nicematrix}
\usepackage{tikz}
\usepackage{xcolor}
\usepackage[
   style=ieee,
 ]{biblatex}

\usepackage{tabularx} 

\usepackage{mathtools}
\usepackage{amsthm}
\theoremstyle{plain} 
\newtheorem{remark}{Remark}
\newcounter{problem}

\newcommand{\linebreakand}{%
  \end{@IEEEauthorhalign}
  \hfill\mbox{}\par
  \mbox{}\hfill\begin{@IEEEauthorhalign}
}
\definecolor{niceblue}{rgb}{0.125, 0.406, 0.852}

\usepackage{float}
\usepackage{bm}
\usepackage{algorithm}
\usepackage{algpseudocode}
\usepackage{booktabs}
\usepackage{multirow}

\usepackage{amsmath,amssymb,amsfonts}
\usepackage{graphicx}
\usepackage{textcomp}
\usepackage{xcolor}
\def\BibTeX{{\rm B\kern-.05em{\sc i\kern-.025em b}\kern-.08em
    T\kern-.1667em\lower.7ex\hbox{E}\kern-.125emX}}

\begin{document}

\title{Fast and Certified Bounding of Security-Constrained DCOPF via Interval Bound Propagation}

\author{\IEEEauthorblockN{Eren Tekeler, Samuel Chevalier}
\IEEEauthorblockA{\textit{Electrical and Biomedical Engineering} \\
\textit{University of Vermont}\\
Burlington, VT, USA \\
\{etekeler, schevali\}@uvm.edu}
\and
\IEEEauthorblockN{Xiangru Zhong, Huan Zhang}
\IEEEauthorblockA{\textit{Electrical and Computer Engineering} \\
\textit{University of Illinois at Urbana-Champaign}\\
Champaign and Urbana, IL, USA \\
\{xiangru4, huanz\}@illinois.edu}

}

\maketitle

\begin{abstract}
Security-Constrained DC Optimal Power Flow (SC-DCOPF) is an important tool for transmission system operators, enabling economically efficient and physically secure dispatch decisions.
Although CPU-based commercial solvers (e.g., Gurobi) can efficiently solve SC-DCOPF problems with a reasonable number of security constraints, their performance degrades rapidly as both system size and the number of contingencies grow into thousands.
In this paper, we design a computational graph representation of the SC-DCOPF-based market-clearing problem, inspired by the third ARPA-E Grid Optimization Competition. Using a tool from the neural network verification community known as Interval Bound Propagation (IBP), we quickly compute bounds on the optimal objective across the full set of N-1 contingencies. Our results demonstrate that IBP can compute certified bounds with mean optimal solution gaps below 3.98\% on small cases, and it can efficiently scale up to 8,316 bus systems with thousands of contingencies. 

\end{abstract}


\section{Introduction}
The most recent Advanced Research Projects Agency-Energy (ARPA-E) Grid Optimization competition (GO3) focused on solving the multi-period Security Constrained Optimal Power Flow problem~\parencite{go3analysis}. The objective of this formulation was to maximize the market surplus while addressing various operational challenges. These challenges involved AC Optimal Power Flow (ACOPF) with soft constraints on nodal power imbalances and line flow violations, unit commitment, topology optimization, reserve constraints, and generator start-up and shut-down constraints. The multi-period aspect of the problem was explored across three different time horizons: real-time, day-ahead, and week-ahead. Furthermore, post-contingency line flows were evaluated using a linearized approximation of ACOPF, known as the DC Optimal Power Flow (DCOPF). 

\begin{figure}[t]
    \centering
    \includegraphics[width=1.0\linewidth]{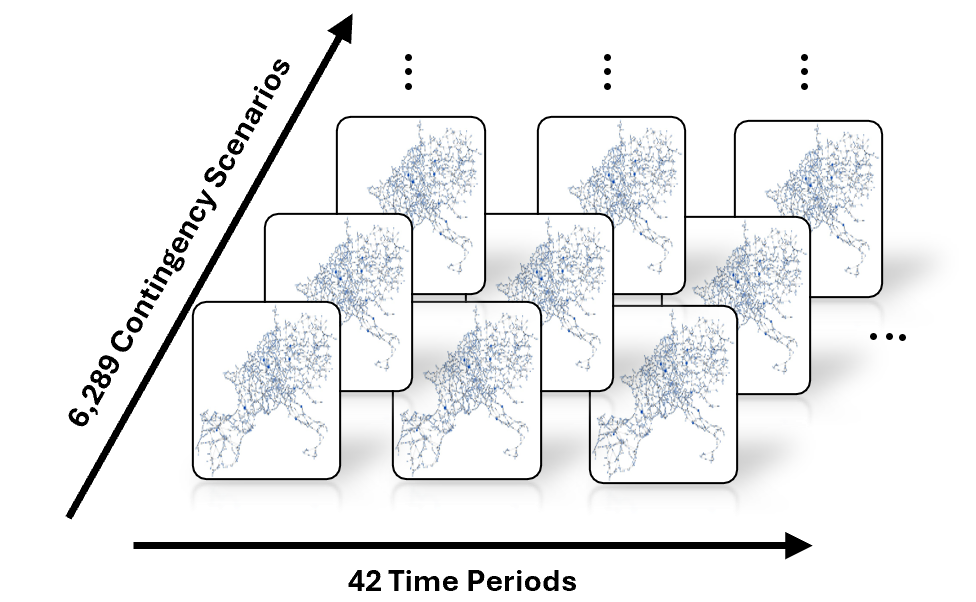}
    \vspace{-5mm}
    \caption{Complexity dimensions of the largest GO3 test case (8,316-bus) solved in this paper.}
    \label{fig: Problem complexity depiction}
    \vspace{-4.5mm}
\end{figure}

The GO3 test instances are designed with multiple dimensions of complexity, where problem size, number of contingencies, and time periods add further layers of difficulty, as illustrated in Fig. \ref{fig: Problem complexity depiction}. Using traditional approaches, the problem requires solving large coupled systems of equations for each contingency scenario and time period, making them computationally challenging. In GO3, competitors predominantly employed decomposition and relaxation-based methods to solve the problems~\parencite{go3analysis}. One competitor, quasigrad, employed a parallelized Adam-based approach on CPUs~\parencite{go3Sam}; however, no team explored a GPU-based strategy that considered the full set of contingencies to address the problem. Recently, authors in \cite{gpuacceleratedSCDOPF} proposed a GPU-based proximal message passing algorithm to address this gap with GPU acceleration. Another approach employed Neural Networks (NNs) to predict solutions to the SC-DCOPF problem with small optimality gaps \cite{deepopf}. In our paper, we focus on formulating the SC-DCOPF market-clearing problem and computing bounds on the objective across the full set of N-1 contingencies. Inspired by the GO3 formulation, we model the security constraints via DC power flow, while the base-case problem is modeled via soft constraints on nodal power imbalances and line flow violations. We enforce hard constraints only for generator limits and on the admissible range of flexible demand.
This naturally enables us to represent the problem as a computational graph with bounded input, which shares a structural similarity with the NN verification problem.

NN verification, which seeks to guarantee or deny the existence of adversarial inputs that violate some output metric over a bounded input, is a hard, nonconvex optimization problem. However, recent advances in GPU computing have enabled scalable and efficient verification techniques, which are broadly categorized as incomplete or complete. Incomplete methods, such as Interval Bound Propagation (IBP)~\parencite{IBP}, compute provable bounds on NN outputs over the entire input domain using interval arithmetic, offering high speed and scalability at the cost of looser bounds. More advanced incomplete approaches, such as CROWN~\parencite{Crown} and $\alpha$-CROWN~\parencite{α-CROWN}, construct convex relaxations to provide provable but not guaranteed global bounds, trading some speed for tighter guarantees. Complete verification methods, such as Branch and Bound \parencite{BaB-NNV}, combine input domain partitioning with incomplete methods to compute provable global bounds given sufficient time.

In this paper, we use IBP to quickly compute relaxed bounds for hard power grid optimization problems. We refer to the resulting bounds as ``certified"~\cite{xu2020automatic}, in the sense that they are guaranteed convex relaxations of the true solution bounds. To compute these bounds, we cast the GO3-inspired SC-DCOPF as a computational graph, enabling our use of NN verification techniques and GPU acceleration. The key contributions are summarized as follows:
\begin{enumerate}
    \item We translate the SC-DCOPF mathematical model into a computational graph in PyTorch, enabling GPU acceleration and compatibility with the NN verification framework $\alpha,\beta$-CROWN.
    \item Leveraging its scalability, computational speed, and tightness, we apply IBP to a range of GO3 test cases.
    \item We present the approach as a tool for identifying infeasible (i.e., welfare-negative) instances: if our conservative upper bound has a negative market surplus,  then the true solution must have a negative market surplus.
\end{enumerate}

In Sec.~\ref{sec: DCOPF}, we walk through the formulation of the DCOPF, with a focus on the GO3-motivated SC-DCOPF model.  We present the computational graph-based bound computation in Sec.~\ref{sec: DCOPF-NN}. We then discuss the results in Sec.~\ref{sec: Results} and conclude with final remarks in Sec.~\ref{sec: Conclusion}.

\section{Problem Formulation}
\label{sec: DCOPF}
In this section, we introduce the standard market-clearing DCOPF problem with flexible demand, and then we define the soft-constrained SC-DCOPF problem.

\subsection{Market Clearing DCOPF}
DCOPF is a linearized approximation of the ACOPF equations, used to solve for optimal generation dispatch in transmission systems. The approximation relies on the assumptions that the voltage angle difference between buses, $\theta_i-\theta_k$, is small (enabling trigonometric terms to be linearized using small-angle approximation), the resistive losses are negligible ($x\gg r$ in transmission lines), and that all bus voltage magnitudes satisfy $V_i\approx1$. The DC power flow mappings are given by
\begin{subequations} \label{eq: DCPF mappings}
\begin{align}
    p_\mathrm{inj, r} &= Y_\mathrm{B,r}\theta_\mathrm{r} \label{eq: pinj} \\
    p_\mathrm{f} &= Y_\mathrm{D}E_\mathrm{r}\theta_\mathrm{r} \label{eq: pf}
\end{align}
\end{subequations}
where the vector $p_\mathrm{inj, r} \in \mathbb{R}^{n_b-1}$ represents the net nodal injections caused by generations and demands, with the row corresponding to the slack bus removed. The reduced nodal admittance matrix $Y_\mathrm{B,r}\in\mathbb{R}^{(n_b-1) \times (n_b-1)}$ maps the reduced voltage angle vector $\theta_\mathrm{r}$ to the net injections. Moreover, the line flow vector, $p_\mathrm{f}\in\mathbb{R}^{n_l}$, is computed from the reduced voltage angle vector $\theta_\mathrm{r}$ through the diagonal line admittance matrix $Y_\mathrm{D}\in\mathbb{R}^{n_l \times n_l}$ and the reduced incidence matrix $E_\mathrm{r}\in\mathbb{R}^{n_l \times (n_b-1)}$. 
To obtain a linear mapping from reduced net nodal injections to line flows, we left multiply \eqref{eq: pinj} with $Y_\mathrm{B, r}^{-1}$ and substitute the term $\theta_r$ into \eqref{eq: pf}. Leveraging the decomposition $Y_\mathrm{B,r}=E_\mathrm{r}^T Y_\mathrm{D} E_\mathrm{r}$, we define the reduced Power Transfer Distribution Factor (PTDF) matrix $\Phi_\mathrm{r}$ as
\begin{equation}
    p_\mathrm{f} = 
    \underbrace{Y_\mathrm{D}E_\mathrm{r}(E_\mathrm{r}^T Y_\mathrm{D} E_\mathrm{r})^{-1}}_{\Phi_\mathrm{r}} \, p_\mathrm{inj, r}.
    \label{eq: line flow ptdf}
\end{equation}
We note that the full PTDF matrix, $\Phi\in \mathbb{R}^{n_l\times n_b}$ is constructed by zero padding the reduced PTDF matrix $\Phi_\mathrm{r}$ to account for the slack bus column as $\Phi = [\mathbf{0} \text{ }\Phi_\mathrm{r}]$. Additionally, we define matrices $N_{\rm g}\in\mathbb{R}^{n_b\times n_g}$ and $N_{\rm d}\in\mathbb{R}^{n_b\times n_d}$ to map the generations and demands to the buses, respectively. Using the generation vector $p_\mathrm{g}\in\mathbb{R}^{n_g}$ and demand vector $p_\mathrm{d}\in\mathbb{R}^{n_d}$, we define the net nodal injection vector as $p_\mathrm{inj} \coloneqq N_{\rm g}p_\mathrm{g}-N_{\rm d}p_\mathrm{d}$.

Using this approximation, we formulate the following DCOPF problem, which
maximizes social welfare by optimally dispatching generation and demand, while respecting power balance, line flow, generation, and demand constraints:\begin{subequations}\label{eq: DCOPF}
\begin{align}
\underset{p_{\mathrm{g}}, p_{\mathrm{d}}}{\max}\quad &\sum_{j=1}^{n_d} c_j(p_{\mathrm{d},j})-\sum_{i=1}^{n_g} g_i(p_{\mathrm{g},i})  \label{DCOPF obj func} \\
{\rm s.t.}\quad & 1^{T}p_{\mathrm{g}}-1^{T}p_{\mathrm{d}}=0 \label{eq: DCOPF pb}\\
& p_{\mathrm{f}}^{\rm min}\leq \Phi p_\mathrm{inj}\leq p_{\mathrm{f}}^{\rm max}\label{eq: ptdf}\\
& p_\mathrm{g}^{\rm min}\leq p_\mathrm{g}\leq p_\mathrm{g}^{\rm max} \label{eq: generator-bounds} \\
& p_\mathrm{d}^{\rm min}\leq p_\mathrm{d}\leq p_\mathrm{d}^{\rm max} \label{eq: demand-bounds}
\end{align}
\end{subequations}

\noindent where $c: \mathbb{R}^{n_d}\rightarrow\mathbb{R}^{n_d}$ and $g: \mathbb{R}^{n_g}\rightarrow\mathbb{R}^{n_g}$ represent vectors of piece-wise linear cost functions, applied element-wise to $p_\mathrm{d}$ and $p_\mathrm{g}$.
For clarity, all upper and lower bounds on the constraints are specified element-wise, and the line flow bounds are symmetric, i.e., $p_{\mathrm{f}}^{\rm min}=-p_{\mathrm{f}}^{\rm max}$. We note that the demand is generally flexible, so we refer to \eqref{eq: DCOPF} as a market-clearing problem, consistent with the GO3 terminology.

\subsection{Formulation of Soft-Constrains}
In the GO3 formulation, rather than enforcing hard constraints on power balance and line flows, as in \eqref{eq: DCOPF}, constraint violations are captured using slack variables and soft constraints. We define the nodal power balance violation constraint as 
\begin{subequations}\label{eq: penalized-power-balance}
\begin{align}
    -s_{\mathrm{inj}} \leq (E^T \Phi - \mathrm{I})&p_\mathrm{inj} \leq s_{\mathrm{inj}} \label{eq: power-balance-range} \\
     s_{\mathrm{inj}} \geq \mathbf{0} \label{eq: power-balance-slack}
\end{align}
\end{subequations}

\noindent where $s^\mathrm{inj}\in\mathbb{R}^{n_{b}}$ is the slack variable that captures the base case power balance violations through the nodal power injections, effectively being the soft-constrained representation of \eqref{eq: DCOPF pb}. 
It 
sums incoming and outgoing flows via $E^T \Phi p_\mathrm{inj}$ to balance the net injections caused by generation and demand, based on Kirchhoff's Current Law. Similarly, the line flow violation soft constraints are given by
\begin{subequations}\label{eq: penalized-flow_violations}
\begin{align}
     \Phi p_\mathrm{inj} - p_\mathrm{f,b}^{\rm max}&  \leq s_{\mathrm{f,b}} \label{eq: power-balance-range} \\
      p_\mathrm{f,b}^{\rm min} -\Phi p_\mathrm{inj}&  \leq s_{\mathrm{f,b}} \\
     s_{\mathrm{f,b}} \geq \mathbf{0}
\end{align}
\end{subequations}

\noindent in which $s_{\mathrm{f,b}}\in\mathbb{R}^{n_l}$ is the slack variable that compensates for the base case line flow violations. Line flows, $\Phi p_\mathrm{inj}$, can take both positive and negative values depending on the flow direction. Thus, flow constraints are expressed symmetrically to account for both upper and lower bounds ($p_\mathrm{f,b}^{\rm min}=-p_\mathrm{f,b}^{\rm max}$), enabling $s_\mathrm{f,b}$ to capture the magnitude of the violation. 

\subsection{Formulation of Security-Constrains}
In addition to the soft-constrained nodal power injection and line flow violations, the GO3 formulation incorporates soft security constraints. This is to ensure system operability under a set of N-1 contingencies by accounting for potential line flow violations in each contingency. However, incorporating all contingencies into the SC-DCOPF problem is computationally challenging, as each contingency requires inverting a new $Y_\mathrm{B,r}$ matrix to obtain an updated contingency PTDF matrix, which could map the net nodal injections to line flows. To overcome this, we model the contingency admittance matrix as
\begin{equation}
    {Y_\mathrm{B,r}^i} = {E_\mathrm{r}^T Y_\mathrm{D} E_\mathrm{r} + e_i y_i e_i^T}.
    \label{eq: nodal admittance rank-1 update}
\end{equation}

\noindent In \eqref{eq: nodal admittance rank-1 update}, $Y_\mathrm{B,r}^i$ represents a nodal admittance matrix with its $i^{\rm th}$ line removed. Via the 
Woodbury matrix identity~\parencite{Horn:1990}, this matrix is computed using a rank-1 update applied to $Y_\mathrm{B,r}$, where $e_i\in\mathbb{R}^{n_b-1}$ is the $i^{\rm th}$ row of $E_\mathrm{r}$ and scalar $y_i$ is the negative of the $i^{\rm th}$ diagonal element of $Y_\mathrm{D}$.  The outer product $e_i y_i e_i^T$ forms a rank-1 matrix that effectively removes the line admittance of the $i^{\rm th}$ line from $Y_\mathrm{B,r}$. By expanding \eqref{eq: line flow ptdf} using the decomposition in \eqref{eq: nodal admittance rank-1 update} and introducing the matrix $M_i$ to cancel the $i^{\text{\rm th}}$ diagonal element of $Y_\mathrm{D}$, the contingency flows can be expressed as
\begin{equation}
    p_\mathrm{f,c}^{i} = M_i Y_\mathrm{D} E_\mathrm{r} (\underbrace{E_\mathrm{r}^T Y_\mathrm{D} E_\mathrm{r} + e_i y_i e_i^T}_{Y_\mathrm{B,r}^i})^{-1} p_\mathrm{inj,r}
    \label{eq: ctg flow computation}
\end{equation}

\noindent where $p_\mathrm{f,c}^{i}$ denotes the contingency flow vector associated with an outage at the line $i$. The $M_i\in\mathbb{R}^{n_l\times n_l}$ is an identity matrix with its $i^{\rm th}$ diagonal element set to zero, effectively removing the corresponding line admittance from $Y_\mathrm{D}$. However, computing \eqref{eq: ctg flow computation} is expensive due to the need for inverse computation for each contingency. To accelerate contingency flow computations, the Sherman–Morrison theorem is applied, leveraging \eqref{eq: nodal admittance rank-1 update} to compute the inverse of the rank-1 updated matrix as
\begin{equation}
    {Y_\mathrm{B,r}^i}^{-1} = Y_\mathrm{B,r}^{-1} - \frac{Y_\mathrm{B,r}^{-1}e_iy_ie_i^T Y_\mathrm{B,r}^{-1}}{1+e_i^TY_\mathrm{B,r}^{-1}e_iy_i}.
\end{equation}
This enables efficient computation of the rank-1 updated matrix (${Y_\mathrm{B,r}^i}$) inverse using only the inverse of the base nodal admittance matrix $Y_\mathrm{B,r}$, avoiding the need to recompute a full matrix inverse for each contingency scenario. Substituting this into \eqref{eq: ctg flow computation} and simplifying, we get
\begin{equation}
    p_\mathrm{f,c}^{i} = M_i \Phi_\mathrm{r} p_\mathrm{inj,r} - M_i Y_\mathrm{D} E_\mathrm{r} u_ig_i(u_i^T p_\mathrm{inj,r}).
    \label{eq: ctg flow simplified}
\end{equation}

\noindent In \eqref{eq: ctg flow simplified}, $u_i\in\mathbb{R}^{n_b-1}$ is defined as $u_i \coloneqq Y_{B,r}^{-1}e_i$ and consequently, $u_i^T=e_i^TY_\mathrm{B,r}^{-1}$ due to the square symmetric nature of the $Y_{B,r}^{-1}$ matrix. Additionally, the scalar term is simplified as $g_i\coloneqq\frac{y_i}{1+e_i^Tu_iy_i}$. To improve computational efficiency by avoiding forming the outer product $u_i u_i^{T}$, 
the second term is decomposed into $M_i Y_\mathrm{D} E_\mathrm{r} u_i g_i$ and $(u_i^{T} p_{\mathrm{inj, r}})$, 
resulting in a vector-scalar product. The constants $u_i$ and $g_i$ are precomputed once for each contingency and stored in memory.

For the set of N-1 contingencies the line flow violations are evaluated similarly to \eqref{eq: penalized-flow_violations} as
\begin{subequations} \label{eq: penalized-ctg-flow_violations}
\begin{align} 
     p_\mathrm{f,c}^{i} - p_\mathrm{f,c}^{\rm max}&  \leq s^i_{\mathrm{f,c}} \\
      p_\mathrm{f,c}^{\rm min} - p_\mathrm{f,c}^{i}&  \leq s^i_{\mathrm{f,c}} \\
     s^i_{\mathrm{f,c}} \geq \mathbf{0}
\end{align}
\end{subequations}
where $s_\mathrm{f,c}^i\in\mathbb{R}^{n_l}$ is the associated slack variable for contingency at line $i$ and the symmetric flow limits are denoted with $p_\mathrm{f,c}^{\rm min}$ and $p_\mathrm{f,c}^{\rm max}$. The total scalar contingency line flow violation is defined as
\begin{equation} \label{eq: scalar-ctg-flow-vio}
    s_\mathrm{f,c}^\mathrm{agg} = \sum^{n_c}_{i=1}{\bm 1}^Ts_\mathrm{f,c}^i
\end{equation}

\noindent where $s_\mathrm{f,c}^\mathrm{agg}\in\mathbb{R}$ represents the aggregated contingency line flow violations across all contingency scenarios, each computed efficiently with the rank-1 update formulation. 

\subsection{Security Constrained DCOPF Formulation}
The GO3 formulation is a multi-period problem by its nature due to time-dependent reserve and generation constraints. In the scope of this paper, intertemporal constraints are neglected (i.e., relaxed), and the problem is modeled as a sequence of independent single-period optimizations. We formulate the modified market-clearing SC-DCOPF objective as 
\begin{equation}
\begin{aligned}
    f(p_{\mathrm{d}}, p_{\mathrm{g}}, s) \coloneqq & \sum_{j=1}^{n_{\mathrm{d}}} c_j(p_{\mathrm{d},j}) - \sum_{i=1}^{n_{\mathrm{g}}} g_i(p_{\mathrm{g},i}) \\
    & - \zeta_{\mathrm{inj}} \mathbf{1}^T s_{\mathrm{inj}} - \zeta_{\mathrm{f}} (\mathbf{1}^T s_{\mathrm{f,b}} + s_{\mathrm{f,c}}^{\mathrm{agg}}).
\end{aligned}
\label{eq: penalized-objective}
\end{equation}
 In \eqref{eq: penalized-objective}, the slack vector $s$ is used as a shorthand for the set $\{s^{\mathrm{inj}}, s_{\mathrm{f,b}}, s_{\mathrm{f,c}}^{\mathrm{agg}}\}$. The penalty coefficients $\zeta_\mathrm{inj}, \zeta_\mathrm{f}\in\mathbb{R}_{++}$ are expressed in dollars per energy ($\$/pu\text{-}h$). They penalize the base-case nodal injection, base-case line flow violations and contingency flow violations, respectively. By incorporating the penalty coefficients for violations, all terms are effectively represented in dollars ($\$$) in the objective function. We state the full formulation of the SC-DCOPF in Problem \ref{problem: SC-DCOPF}.
\begin{center}  
\vspace{2pt}
\begin{minipage}{\columnwidth}
\hrule
\vspace{2pt}
\refstepcounter{problem}
\noindent \textbf{Problem 1:} Soft Constrained SC-DCOPF Problem  \label{problem: SC-DCOPF}
\vspace{2pt}
\hrule
\begin{subequations}
\begin{align} 
\max_{\substack{p_\mathrm{g}, p_\mathrm{d}, s}} &\quad f(p_{\mathrm{d}}, p_{\mathrm{g}}, s)\\
    {\rm s.t.}\; &\quad \text{Hard constraints: } \eqref{eq: generator-bounds}, \eqref{eq: demand-bounds}\\ 
    &\quad \text{Soft constraints: } \eqref{eq: penalized-power-balance}, \eqref{eq: penalized-flow_violations}, \eqref{eq: penalized-ctg-flow_violations}, \eqref{eq: scalar-ctg-flow-vio} 
\end{align}
\end{subequations}
\vspace{-10pt} 
\hrule
\end{minipage}
\vspace{2pt}
\end{center}

\section{Graph-Based Bound Computation}
In this section, we first describe the computational graph formulation of the SC-DCOPF problem. Then, we briefly present the NN verification and IBP.
\label{sec: DCOPF-NN}

\subsection{Computational Graph Formulation}
Computational graphs represent the flow of operations from input to output, enabling a structured way to perform forward and backward passes. They are at the heart of GPU-accelerated bound propagation tools, such as $\alpha,\beta$-CROWN to compute bounds. Thus, we cast the Problem \ref{problem: SC-DCOPF} as a computational graph to leverage these verification frameworks. 
\begin{remark} \label{remark: equality-argument}
The non-negative penalty coefficients in \eqref{eq: penalized-objective} ensure that the optimizer minimizes the slack variables to their smallest feasible values. Thus, at optimality, each slack variable is exactly equal to the respective violation magnitude. 
\end{remark}
Leveraging Remark \ref{remark: equality-argument}, we model the soft constraints of Problem \ref{problem: SC-DCOPF} as deterministic equations. The nodal power imbalance constraints in \eqref{eq: penalized-power-balance} are represented as
\begin{equation}
    s_{\mathrm{inj}} = \left| (E^T \Phi - \mathrm{I})p_{\rm inj} \right| \label{eq: slack-sinj}
\end{equation}
where the intermediate expression $s_\mathrm{inj}$ captures the nodal power imbalance violations. Similarly, the base case line flow violations in \eqref{eq: penalized-flow_violations} can be formulated as
\begin{equation}
    s_{\mathrm{f,b}} = \max\left( \left| {\Phi p_\mathrm{inj}} \right| - p_\mathrm{f,b}^{\rm max} ,0\right) \label{eq: slack-sfb}
\end{equation}
in which the $\max(\cdot)$ operator is applied element-wise and $s_{\mathrm{f,b}}$ quantifies any flow exceeding the limit $p_\mathrm{f,b}^{\rm max}$. Additionally, for the GO3 test instances with hundreds or thousands of contingencies, a loop-free model implementation is essential for traceability of the computational graph and faster computation times. With pre-processing, \eqref{eq: ctg flow simplified} is reformulated as
\begin{equation}
    P_\mathrm{f,c} = M\otimes(\mathbf{1}(\Phi p_\mathrm{inj})^T) - B\otimes((Up_\mathrm{inj})\mathbf{1}^T).
    \label{eq: ctg flow matrix form}
\end{equation}

\noindent $M \in \mathbb{R}^{n_\mathrm{ctg} \times n_l}$ is the preprocessed contingency matrix, with each row being all ones except for the contingency line entry, which is zero. The outer product $\mathbf{1}(\Phi p_\mathrm{inj})^T$ replicates the line flow vector, \( \Phi p_{\mathrm{inj}} \), across rows to match the number of contingencies. The element-wise (Hadamard) product, denoted by $\otimes$, sets the line flow corresponding to the contingency to zero. With a slight simplification, $M_i Y_\mathrm{D} E_\mathrm{r} u_ig_i$ vector can be denoted as $b_i$ and can be stacked as $B\coloneqq[b_1 \text{ } b_2 \text{ } b_3 \text{ } \dots \text{ } b_{n_\mathrm{ctg}}]^T$, where $B\in\mathbb{R}^{n_\mathrm{ctg}\times n_{l}}$. Additionally, using the full $p_\mathrm{inj}$ vector requires zero padding for $u_i$ when building the $U$ matrix. The padded vector $u'_i\coloneqq[0 \text{ } {u_i}^T]^T$ forms the matrix $U\coloneqq[u'_1 \text{ } u'_2 \text{ } u'_3 \text{ } \dots \text{ } u'_{n_\mathrm{ctg}}]^T$, where $U\in\mathbb{R}^{n_\mathrm{ctg}\times n_{b}}$. This avoids explicit looping over all contingencies and eliminates the need to store each $M_i$ separately, resulting in significant computational efficiency. Similar to \eqref{eq: slack-sfb}, $S_\mathrm{f,c}\in\mathbb{R}^{n_\mathrm{ctg} \times n_l}$ denotes the contingency line flow expression, which computes the violations as
\begin{equation}
        S_{\mathrm{f,c}} = \max\left( \left|P_\mathrm{f,c}\right| - \mathbf{1} \left(p_\mathrm{f,c}^{\rm max}\right)^T ,0\right). \label{eq: slack-Sc}
\end{equation}
In \eqref{eq: slack-Sc}, the line flow violations are captured for all contingencies. The total violation is then obtained as
\begin{equation}
    s_\mathrm{f,c}^\mathrm{agg} = \mathbf{1}^TS_{\mathrm{f,c}}\mathbf{1} \label{eq: slack-sagg}
\end{equation}
where $s_\mathrm{f,c}^\mathrm{agg}$ represents the aggregated contingency line flow violations across all N-1 contingencies.

\subsection{Cost Function Implementation}
In the GO3 formulation, the cost functions for generators and demands are modeled as convex and concave piece-wise linear functions, respectively. Given a piece-wise linear curve whose $i^{\rm th}$ segment has slope $a_i$ and horizontal shift $b_i$, the curve can be represented using ReLU activations via \eqref{eq: combined ReLUs}:
%
%
\begin{subequations}
    \begin{align}
    &F_i(x) = a_i \, \mathrm{ReLU}(x - b_i) \label{eq: single ReLU} \\
    &F_i^*(x) = \min(F_{i-1}^*(x) + F_i(x), U_i) \label{eq: combined ReLUs}
    \end{align}
\end{subequations}
where $U_i$ is the cumulative upper bound of segment $i \in \{1,2,\dots, n_{\rm seg}\}$, and $F_{0}^*(x)=0$. We apply slope scaling and horizontal shift for each segment via \eqref{eq: single ReLU}, while in \eqref{eq: combined ReLUs}, individual ReLUs are combined through addition and clipped to enforce the piecewise limits.

\subsection{Interval Bound Propagation}
In this paper, we seek to use IBP to quickly bound the solution to Problem~\ref{problem: SC-DCOPF}. Customarily, IBP efficiently propagates bounded inputs through NN layers in NN verification problems (see next subsection) to conservatively compute element-wise output bounds. Let the layer input $x\in\mathbb{R}^n$ be bounded by $\ell_{\infty}$ norm with a positive element-wise perturbation margin $\epsilon$. The interval that  $x$ lies in is defined as $\underline{x}\leq x \leq \bar{x}$, where $\underline{x}=x-\epsilon$ and $\bar{x}=x+\epsilon$ are lower and upper bounds.  Defining $x_\mu\coloneqq\frac{1}{2}(\bar{x}+\underline{x})$ and $x_\sigma\coloneqq\frac{1}{2}(\bar{x}-\underline{x})$, the output bounds can be computed for the NN layer $y=\sigma(Wx+b)$ as
\begin{subequations}\label{eq: IBP}
   \begin{align}
    &\bar{y}=\sigma(|W|x_\sigma + Wx_\mu + b) \\
    &\underline{y}=\sigma(-|W|x_\sigma + Wx_\mu + b) 
    \end{align} 
\end{subequations}
where $\sigma(\cdot)$ is any element-wise applied monotonic non-decreasing activation function \parencite{IBP}. IBP computes \eqref{eq: IBP} for each layer and propagates bounds to achieve NN output bounds. The bounds $\bar{y}$ and $\underline{y}$ are ``certified", meaning there exists no $x\in[\underline{x}, \bar{x}]$ such that $y\notin [\underline{y},\bar{y}]$. Generally, IBP will loosely bound the true reachable set.

\subsection{Formal Verification of Neural Networks}
\label{sec: NN verification}
The NN verification problem aims to formally prove that an NN satisfies a given property over a bounded input. Let $f(x): \mathbb{R}^n\rightarrow\mathbb{R}^m$ denote a NN mapping, and the input set is defined by $\mathcal{C}=\{ x\in\mathbb{R}^n \mid \|x - x_0\|_p \leq \epsilon \}$ where $x_0$ defines the center and $\epsilon$ is the perturbation radius. We first identify the worst-case output by posing:
\begin{equation}
y^*\coloneqq\min_{x \in \mathcal{C}} \; c^Tf(x)
\label{NN Verification Objective}
\end{equation}
where $c\in\mathbb{R}^m$ defines the property to be verified on the output $f(x)$ and $y^*$ denotes the worst-case value of this property. The second stage of the problem checks the satisfaction of the following inequality:
\begin{equation}
    y^*>\gamma \label{eq: NN verification}
\end{equation}
where $\gamma$ is a scalar specification term. If \eqref{eq: NN verification} is satisfied, we call the property verified, indicating that the satisfaction of the property for all $x\in\mathcal{C}$. Since NN mappings are generally non-convex, solving \eqref{NN Verification Objective} optimally is a hard problem. Thus, instead of solving it, bound propagation techniques are employed to quickly compute bounds on $y^*$. 

Problem \ref{problem: SC-DCOPF} shares a similar structure with \eqref{NN Verification Objective}. As stated in Remark \ref{remark: equality-argument}, the variables that $s$ represent in Problem \ref{problem: SC-DCOPF} are no longer variables but intermediate expressions, which leaves the problem only with bounded inputs $p_g$ and $p_d$. Similar to the affine layer and nonlinear activation interactions in NNs, the SC-DCOPF graph is constructed through sequential operations, where linear mappings interact with nonlinear operators during cost and violation computations. This similarity allows the problem to be formulated analogously to an NN verification task, enabling the bound computations with GPU-assisted IBP.

We are interested in two applications on the SC-DCOPF problem: $(i)$ computing bounds on the optimal solution via IBP, and assessing tightness and computation time, $(ii)$ posing the NN verification problem to identify infeasible or welfare-negative instances. A case is classified as infeasible or welfare-negative when the following inequality holds:
\begin{equation}
    \max_{x \in \mathcal{C}} f(x)<0 \label{eq: inf-or-wn}
\end{equation}
where $f(x)$ is a placeholder for the SC-DCOPF computational graph and $\mathcal{C}$ defines the set of feasible generations and demand. When \eqref{eq: inf-or-wn} is satisfied, it implies that no input within the set yields a non-negative objective, indicating either constraint violations, negative social welfare, or both.

\section{Results}
\label{sec: Results}
In this section, we present simulated test results for the bounding performance of IBP on the GO3 competition instances. In the appendix (see \eqref{eq: f-ibp}), we state the full, explicit computational graph that we apply IBP to.

\subsection{Test Setup}
We formulated\footnote{The implementation of the proposed method is publicly available \cite{project_repo}.} the Problem \ref{problem: SC-DCOPF} in Julia using JuMP and employed Gurobi 13 \cite{gurobi} for benchmarking. Computational graphs were built in PyTorch, and IBP computations were performed using the $\alpha,\beta$-CROWN framework. All experiments were performed on the Vermont Advanced Computing Center (VACC) using AMD EPYC 7763 CPUs and NVIDIA H200 GPUs. We collected a total of 275 test instances from different events and divisions of the GO3 competition.

\begin{figure}[t]
    \centering
    \includegraphics[width=1\linewidth]{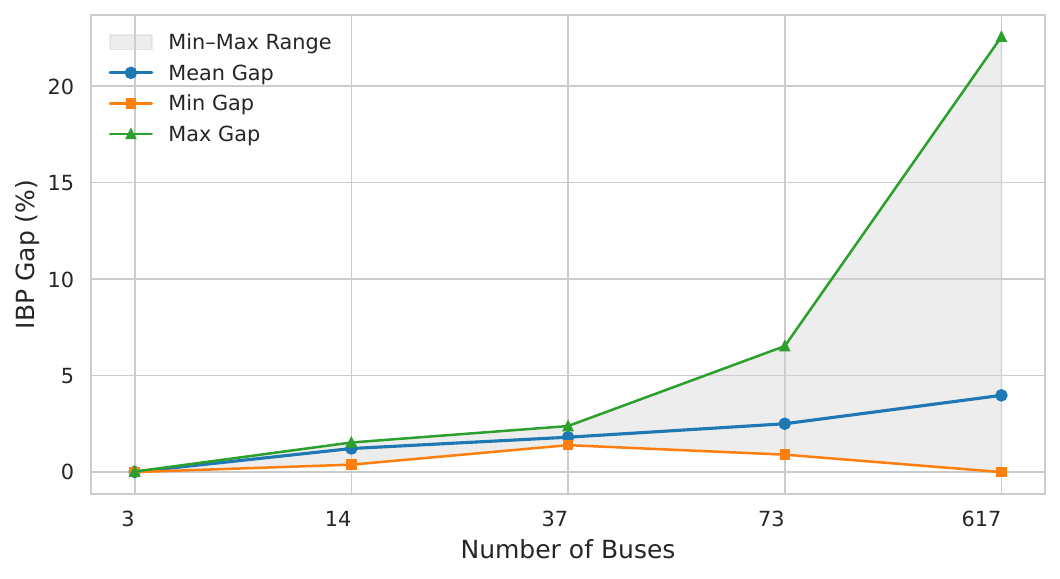}
    \vspace{-5mm}
    \caption{IBP gap statistics across different benchmarked system sizes. Mean gaps remained below 3.98\% for all system sizes.}
    \label{fig:IBP gap}
    \vspace{-4.5mm}
\end{figure}

\begin{table}[h]
\centering
\caption{Breakdown of the Test Cases}
\label{tab:test_system_sizes}
\footnotesize
\setlength{\tabcolsep}{4pt}
\begin{tabular}{ccccc}
\hline
\textbf{\#Buses} & 
\textbf{\#Contingencies} & 
\textbf{\#Devices} & 
\textbf{\#Lines} & 
\textbf{\#Time Indices} \rule{0pt}{2.3ex} \\
\hline
3    & 2        & 3              & 4              & 48,42,18 \\
14   & 12       & 17             & 20,37          & 48,18,42 \\
37   & 40       & 34             & 57             & 48,42,18 \\
73   & 2        & 205            & 120            & 48,42,18 \\
617  & 562      & 499            & 853,975        & 48,18 \\
1576 & 147,219  & 2064,2066      & 2371,2427      & 48,42,18 \\
2000 & 2756     & 1894           & 3206           & 48,18,42 \\
4224 & 2313     & 2151           & 4930           & 48,18,42 \\
6049 & 3884,3902& 3774           & 8006           & 42,18 \\
6717 & 2670     & 5826           & 9140           & 18 \\
8316 & 6289     & 5585,5589,5583 & 11972          & 48,42,18 \\
\hline
\end{tabular}
\end{table}
\noindent The breakdown of the used test cases is presented in the Table \ref{tab:test_system_sizes}. We analyzed 25 instances for each bus size having various numbers of contingencies, devices, lines, and time indices. We were able to obtain the optimal solutions for cases up to 617 bus systems, which served as a benchmark for the tightness analysis of the IBP bounds.
Specifically, we compare the bound provided by IBP with the optimal LP solution computed by Gurobi. We refer to the difference between these solutions as a ``gap":
\begin{align}
{\rm gap}\triangleq\frac{\left|{\rm IBP}-{\rm Gurobi}\right|}{\left|{\rm Gurobi}\right|}.
\end{align}

\noindent Also, IBP bounds were batch computed for all time indices for up to 6717 bus instances, whereas we serially computed the bounds for 8316 bus instances due to memory limitations. 

\subsection{Numerical Results}
We tested IBP in terms of the bound tightness and scalability. Fig. \ref{fig:IBP gap} illustrates the optimality gap statistics across the benchmarked system sizes. For each system size, we solved 25 test instances using Gurobi and computed bounds using IBP. The mean gap remained below 3.98\% for all systems, with the maximum gap occurring in a 617-bus instance at 22.56\%. Additionally, the upper bounds computed in some\footnote{None of the 3, 14, 37, 73, 617 and 4224 bus systems were found to be guaranteed infeasible or welfare-negative. However, 1576, 2000, 6049, 6717, 8316 were found to be 20/25, 25/25, 7/25, 25/25, 12/25 guaranteed infeasible or welfare-negative for at least one time index, respectively.} test instances were negative, implying that for the given input set, the instance is guaranteed to be welfare-negative (i.e., infeasible by GO3 modeling standards). Identifying these cases via IBP can serve as a screening mechanism before solving a full problem, like optimal line switching, and it can streamline computational workflows by eliminating the need to solve for the optimal dispatch.

\begin{table}[h]
\centering
\renewcommand{\arraystretch}{1} 
\caption{IBP Speedup Summary for Benchmarked System Sizes}
\begin{tabular}{c | c c c}
\toprule
\multirow{2}{*}{\textbf{Number of Buses}} & \multicolumn{3}{c}{\textbf{Speedup} ($\times$)} \\
\cmidrule(lr){2-4}
 & \textbf{Min} & \textbf{Mean} & \textbf{Max} \\
\midrule
3    & 1.02      & 2.39     & 3.35      \\
14   & 1.31      & 2.86     & 4.24      \\
37   & 5.23      & 10.01    & 10.94     \\
73   & 7         & 11.14    & 12.25     \\
617  & 5.21k     & 15.07k    & 32.19k    \\
\bottomrule
\end{tabular}
\label{tab:speedup_summary}
\end{table}

\begin{figure}[t]
    \centering
    \includegraphics[width=1\linewidth]{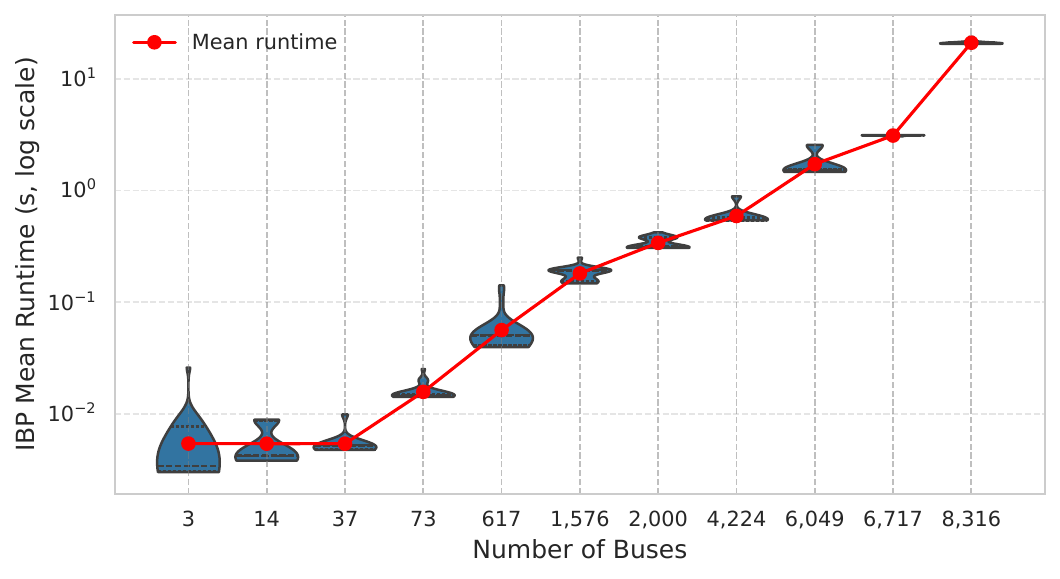}
    \vspace{-5mm}
    \caption{Distribution of IBP mean runtimes in log scale across different system sizes.}
    \label{fig:IBP runtime}
    \vspace{-4.5mm}
\end{figure}

\noindent  Table \ref{tab:speedup_summary} presents statistics on the speedup achieved in bound computations. As the system size grows, IBP becomes increasingly advantageous, achieving speedups of up to 32.19 thousand times. The reported speedups can be interpreted as computational time savings achieved by IBP, eliminating the need to solve infeasible or negative-welfare instances. When including the full set of contingencies, instances beyond 617 buses became challenging for Gurobi due to memory limitations; in contrast, IBP scales efficiently, computing the bounds for systems up to 8,316 buses. Fig. \ref{fig:IBP runtime} shows the distribution of runtimes to compute the bounds of a single time index across multiple test cases. Runtimes scaled efficiently, from a 3-bus system with 2 contingencies to an 8,316-bus system with 6,289 contingencies, with mean runtimes reaching 23.8 seconds. These results showcased the scalability and bound quality of our methodology. 

\section{Conclusion}
\label{sec: Conclusion}
In this paper, we formulated the GO3-inspired SC-DCOPF problem as a computational graph to leverage IBP for GPU-accelerated bound computations. By employing $\alpha,\beta$-CROWN, we evaluated IBP for its bound tightness, scalability, and ability to identify infeasible or negative-welfare instances. Our results demonstrate that it efficiently computes tight bounds and scales to one of the largest instances (8,316 buses with 6,289 contingencies) of the GO3 competition. Future work will investigate extending IBP within Branch and Bound frameworks to leverage its speed and scalability for computing global bounds on challenging power system problems.

\appendices
{\section{}\label{AppA}}
The explicit function to which we apply IBP is given by 
\begin{align} \label{eq:ibp_f}
f_{{\rm ibp}}(p_{\mathrm{d}},p_{\mathrm{g}}) \coloneqq & \sum_{j=1}^{n_{\mathrm{d}}} c_j(p_{\mathrm{d},j}) - \sum_{i=1}^{n_{\mathrm{g}}} g_i(p_{\mathrm{g},i}) \nonumber \\
    & - \zeta_{\mathrm{inj}} \mathbf{1}^T s_{\mathrm{inj}} - \zeta_{\mathrm{f}} (\mathbf{1}^T s_{\mathrm{f,b}} + s_{\mathrm{f,c}}^{\mathrm{agg}}),
\end{align}
where $s_{\rm inj}$, $s_{\mathrm{f,b}}$, and $s_{\mathrm{f,c}}^{\mathrm{agg}}$ are no longer slack variables, but intermediate expressions defined in \eqref{eq: slack-sinj}, \eqref{eq: slack-sfb}, and \eqref{eq: slack-sagg}. IBP then computes an upper bound on the following problem:
\begin{align}\label{eq: f-ibp}
\underset{\substack{p_{\mathrm{g}}^{{\rm min}}\leq p_{\mathrm{g}}\leq p_{\mathrm{g}}^{{\rm max}}\\
p_{\mathrm{d}}^{{\rm min}}\leq p_{\mathrm{d}}\leq p_{\mathrm{d}}^{{\rm max}}
}
}{\max}\quad f_{{\rm ibp}}(p_{\mathrm{g}},p_{d})
\end{align}
where the only bounded variables are $p_{\mathrm{g}}$ and $p_{\mathrm{d}}$.

\section*{AI Usage Disclosure}
Generative AI tools were used to improve the language and clarity of this manuscript. Formulations, analyses, and result population were performed by the author(s), who take full responsibility for the accuracy and integrity of the work. 

\printbibliography

\end{document}